# Exploration of Embodied Space Experience through Umbilical Interaction: A Grounded Theory Approach

Shuai Guo
Department of Informatics and Media
Uppsala University
Uppsala, Sweden
shuai.guo.5671@student.uu.se

Dawei Liu
Department of Informatics and Media
Uppsala University
Uppsala, Sweden
dawei.liu.7239@student.uu.se

Tiantian Zheng
Department of Informatics and Media
Uppsala University
Uppsala, Sweden
tiantian.zheng.6212@student.uu.se

**ABSTRACT**

This paper critiques the limits of human-centered design in HCI, proposing a shift toward Interface-Centered Design. Drawing on Hookway's philosophy of interfaces, phenomenology, and embodied interaction, we created Umbilink, an umbilical interaction device simulating a uterine environment with tactile sensors and rhythmic feedback to induce a pre-subjectivized state of sensory reduction. Participants' experiences were captured through semi-structured interviews and analyzed with grounded theory. Our contributions are: (1) introducing the novel interface type of Umbilical Interaction; (2) demonstrating the cognitive value of materialized interfaces in a human-interface-environment relation; (3) highlighting the design role of wearing rituals as liminal experiences. As a pilot study, this design suggests imaginative applications in healing, meditation, and sleep, while offering a speculative tool for future interface research.

CONCEPTS • Interactive systems and tools • Scenario-based design • Interface design prototyping

## 1 Introduction

What is interface? Broadly speaking, we regard it as an intermediary layer between two systems, serving as a medium for the exchange of information and matter. Hookway traces the origins of the term 'interface' back to its definition in fluid dynamics, as well as evolution of information theory and cybernetics. He summarizes the interface as 'a form of technical relationship' rather than the technology itself [1]. Another important interpretation comes from Grudin, who defined the position of the interface at different historical stages in the retrospective analysis of interface history. The shift in position extends from hardware, software to higher-level cognitive activities. In an analogy with the developmental stages of infancy, the continuity and fluidity of the interface as an abstract concept are revealed [2]. The introduction of phenomenology has provided a new research paradigm for human-computer interaction. Within Merleau-Ponty's phenomenological framework, the body is reoriented from a biological and physical object to a 'lived body,' connecting human consciousness with the external and primordial phenomenal world [3]. The concept of the interface, as an abstract relational form, begins to shift from an external, disembodied space toward the human body.

Currently, academic exploration of human interaction possibilities has covered almost all sensory and motor parts, including skin and haptics, e.g., modulating joint flexibility by modulating skin thermal sensing [4]; leg movements, e.g., providing swinging degrees of freedom to avoid VR vertigo [5], as well as emerging explorations in areas such as the face, for which discriminable normal forces are provided [6]; and the tongue, for scenarios such as healthcare, exploitingh haptic input capabilities [7], to name just a few. Across all interface types, we are keenly aware that the interface itself remains, as Hookway describes its nature, 'ubiquitous and hidden from view' [8], and it is difficult to define the exact location of the interface during these interactions, especially given the separation of input and The separation of the input and output devices is in line with the consistent abstract nature of interfaces. This is certainly not a problem for users, who can use it without any notion of interface. For designers, making interfaces disappear is a goal recognised by the new HCI paradigm, and interfaces should be naturally embedded in life, as Weiser envisions, a 'ubiquitous computing' system [9]. Microsoft Research noted in 2004 that interaction technologies are moving from centralised to distributed, while input and output devices, computation and feedback located in discrete states, shaping the modern human-centred computing environment [10]. However, as researchers, we try to bring the research centre back to the nature of interfaces in our microscopic view, based on the recognition of the past trend of 'ubiquitous computing', we try to reflect on the question: is it possible that in some specific interaction scenarios, we need to make interfaces visible and prominent, in order to discover or perceive its role as a mediator in the exchange of information and matter?

At the same time, we perceive a research gap that is being echoed. While the design of interfaces in current HCI research often revolves around the extension and augmentation of intrinsic perceptual channels, there has been little discussion of the input and output channels common to all humans and other placental mammals—the umbilical cord, the biological interface we are all born with. This imagery appears frequently in art installations and films, intertwined with feminist, paternity and identity-political metaphors of bondage and constraint. Here, by placing the physical properties of the interface at the forefront of mediation, we solidify abstract concepts such as "middle layer" or "relational form" into a balanced, symmetrical and striking entity, whose visual representational properties, such as shape, colour and surface texture, are distinct from its connecting ends. To some extent, this study questions the value of human-centred design concepts for HCI researchers themselves, which is the first topic of this study:

RQ1: We envisage a return of design focus in a particular context, i.e. From 'Human-Centered Design' to 'Interface-Centered Design'. Is there any potential inspiration for HCI researchers?

Our vision is to provide the HCI field with a testing ground for the question "What is signified when we talk about interfaces? The goal of this research is to promote wider philosophical speculatoin about it, to produce knowledge within, or simply adding more design inspiration for the future. Further, in using the research methodology of grounded theory, we are simultaneously reflecting on Hookway's philosophical articulation: the relational form of the interface connects humans and the environment, but also separates them in a way that makes their existence clear; In an attempt to dissolve the differences between humans and their environment, it is maintained in a high-profile manner [8].

In the design of this study, we created an experimental device called the 'Umbilink', a morphological analogue of the umbilical cord, which is connected at one end to a confined environment symbolising the mother's body, and at the other end to the user's navel. We aim to return users to a 'pre-subjectified' state where sensory development is incomplete. We challenge the default assumption in interface design that users are independent subjects. As users imagine themselves returning to the womb, they gradually transition from being absolute subjects in need of service and attention, to 'connection points' that simultaneously possess both subject and object attributes. Thus, we hypothesise that,

RQ2: When the user "slips" from an absolute interacting subject to a part of the interaction loop, is it possible to perceive new ways of bodily integration in the simulation of sensory degradation, and in this way to generate new interaction experience?

The research follows a phenomenological paradigm, where grounded theory is used as a method to explore the subjective structure of the user, but we do not presuppose any possible positions or norms. Because of the lack of cultural knowledge and direct experience with the umbilical cord, we first learnt about the physiological structure of the umbilical cord and cultural metaphors from an ethnographic perspective.

## 2. BACKGROUND

### 2.1 Structure and Symbolism of Umbilical Cord

The umbilical cord is the connecting structure between the embryo and the mother in all placental mammals during the embryonic stage. The connecting stalk forms during the third week of embryonic development and matures by the seventh week. It typically consists of two umbilical arteries and one umbilical vein, all enclosed in a gel-like substance known as Wharton's jelly. The human umbilical cord measures approximately 55 centimeters in length and 2 centimeters in diameter at full term, it is typically cut when the newborn can breathe independently [11]. As an interface between organisms, it connects the mother to the embryo. The umbilical vein supplies oxygen-rich blood and nutrients to the fetus, while the umbilical artery transports carbon dioxide and metabolic waste back to the mother [12], representing the most fundamental input and output process in human life.

We particularly note the metaphorical significance of the umbilical cord across different cultural perspectives. The background research beginning with an anthropological study of how newborn caregivers in central Uganda perceive the umbilical cord. This study was based on the close contact experiences between caregivers and new mothers, as well as the latter's expressions in focus groups, revealing psychological activities under sensitive descriptions. This confirmed our preconception that the umbilical cord is at the center of the birthing process, such as 'the umbilical cord is the child's life' and 'we view the umbilical cord as the center of life.' It also offers novel perspectives that can guide and reflect on design, such as mothers' physiological fear of the umbilical cord's unusual appearance, 'it looks like intestines'; healthcare workers' discomfort stemming from the professional pressure to prevent diseases like neonatal tetanus from entering the newborn's body through the air; and the special value of the umbilical cord in a patrilineal clan-dominated society: a popular superstition serving as a primitive alternative to DNA paternity testing [13]; in the Chennai region of India, the umbilical cord is transformed into an amulet to protect the child's health, symbolizing rebirth [14]; and in Indonesia, the ritual of burying the umbilical cord symbolizes the transmission of life across generations [15]. We have identified numerous value-sensitive points for umbilical interaction design, such as attachment, protection, and separation. By clarifying its symbolic significance, we can encode and analyze users' oral feedback from an anthropological perspective across different regional cultures and explore the role of cultural cognition in creating a new interaction interface.

**2.2 Grounded Theory and Subjective Experience**

The classic Grounded Theory was proposed by Glaser and Strauss in 1967. They provided a scientific approach for sociological research on how to analyse data to arrive at facts. Specifically, it rejected the preconceived 'hypothesis-deduction' approach and instead used a systematic three-level coding method to to generate knowledge from diverse subjective data [16]. Charmaz's constructivist grounded theory reflects on the traditional overemphasis on objectivity and 'seeking meaning,' emphasising the inseparability of the researcher's subjectivity from the data and further elucidating the importance of co-constructing meaning [17]. This introduces a narrative and reflective practical framework to grounded theory, making it more suitable for exploring subjective experiences. Benner, in her research on nursing science, pioneered the integration of grounded theory with phenomenology and introduced studies of various natural groups. She used 'thick description' to describe how 'pain' affects people as a way of giving meaning to the body, providing a model for subsequent analyses of subjective experiences in the HCI field [18] and widely referenced in interface development and user needs exploration.

**3.  prototype DESIGN**

The design of this study is based on short-term prototype design practices in a studio setting, with team members conducting daily regular discussions over a two-week period. We integrated the knowledge of different team members in Arduino hardware development, phenomenology, and media design, focusing on portability, mobility, and the distinctiveness of the interface as the core standards for the prototype design.

**3.1 Prototype Composition**

By reimagining the imagery of the 'uterus-umbilical cord-mother' through interactive reconstruction, we aim to provide users with an experiential environment representing the primal setting of the human habitat. The final design is an interactive device system named 'Umbilink,' which integrates bodily perception with spatial metaphor, comprising an accessible interactive space and a wearable touch-feedback device.

We have created a closed, secure, and private environment that provides users with sensory deprivation and physical protection within the space. The interior features dim, controllable lighting to simulate the experience in the womb, enhancing immersion. The tactile device is built on the Arduino development platform and consists of a microcontroller mainboard, touch sensors, button sensors, a power module, and a vibration and LED system. The core circuit components are encapsulated in a simple waist pouch, with the touch sensor positioned at the central front area of the pouch, aiming to guide users to engage in sensory interaction through the action of 'touching the navel' during the interaction process. A red gift ribbon tightly wraps around the internal 50cm dupont wire, forming a protective layer mimicking Wharton's jelly. The choice of red and the ribbon's undulating motion during movement symbolise the flow of blood between the mother and foetus, visually emphasising its symbolic meaning of 'life connection.' The portable power bank is pre-placed in the participant's pocket as the system's hardware control. During the Arduino IDE coding process, we selected the Finite State Machine (FSM) model as the core architecture, which is suitable for handling discrete processes with clear stages, aligning with the sensor's trigger timing in Umbilink.

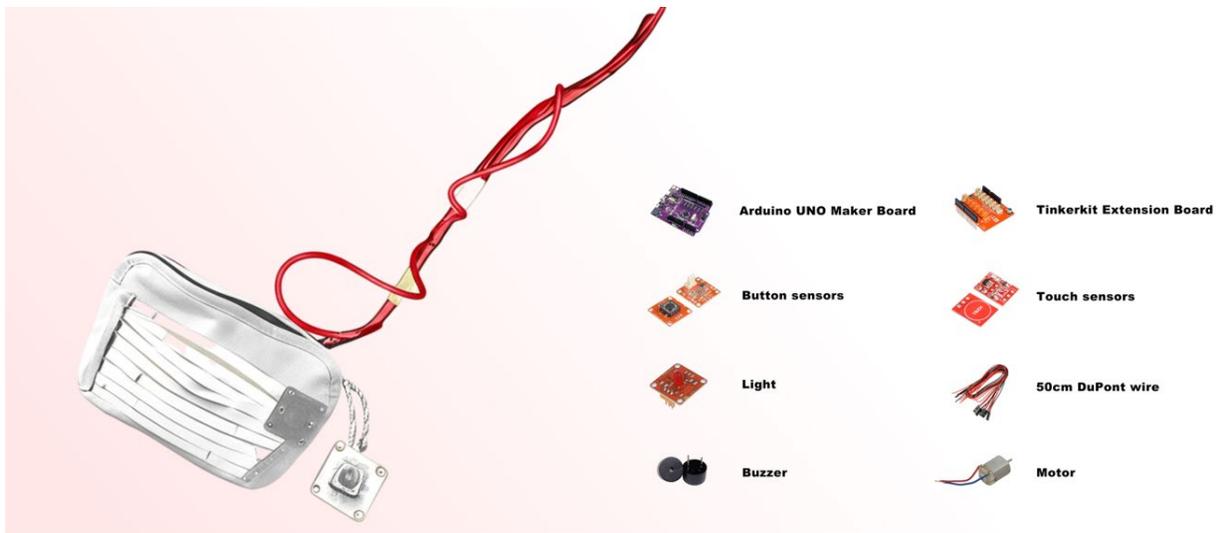

Figure 1: Appearance of Umbilink, components are packaged in a belt bag

**3.2 Interaction Process**

The user experience process is designed as a multi-stage body-space interaction process that relies on the regulation of sound and touch to guide internal rhythms. This includes:

**Startup Phase:** Before the user's interactive experience, we inform them of the recommended posture (cross-legged sitting), the estimated duration (approximately 10–15 minutes), and the conditions for state transitions (pressing the button and approaching the touch sensor), along with precautions and reasons for corresponding actions (e.g., avoid gripping the touch sensor tightly to prevent electromagnetic field disruption). Upon entering the interactive space, one end of the umbilical cord is connected to an Arduino circuit board, while the other end is fitted with an LED light, suspended from the ceiling of the healing space, symbolising the position of the placenta within the uterus. Through the device's structural design within the confined space, a physical and symbolic connection is established between the user and the space.

**Regulation and Feedback Phase:** In its initial state, Umbilink continuously emits low-frequency, rhythmic, and rapid sounds accompanied by synchronised vibrations, serving as the foundational rhythm to guide users in regulating their physical and mental states. Each time the user touches the sensor located on the navel, the time interval between two rhythms increases by 0.2 seconds.

Users may choose to repeatedly touch the sensor during the process to find a frequency most suitable for their current physical sensations. This frequency remains constant until the next touch triggers a new adjustment.

**Exit and Loop Phase:** When the user has touched the sensor 15 times, the system automatically determines that a complete experience cycle has ended. At this point, the LED light flashes again and plays a continuous ascending melody composed of multiple notes, signifying the completion of the experience. The user may choose to exit the interactive space or restart a new experience by clicking the button again. The entire process supports individual rhythm regulation, encouraging users to develop awareness and control over their physical state.

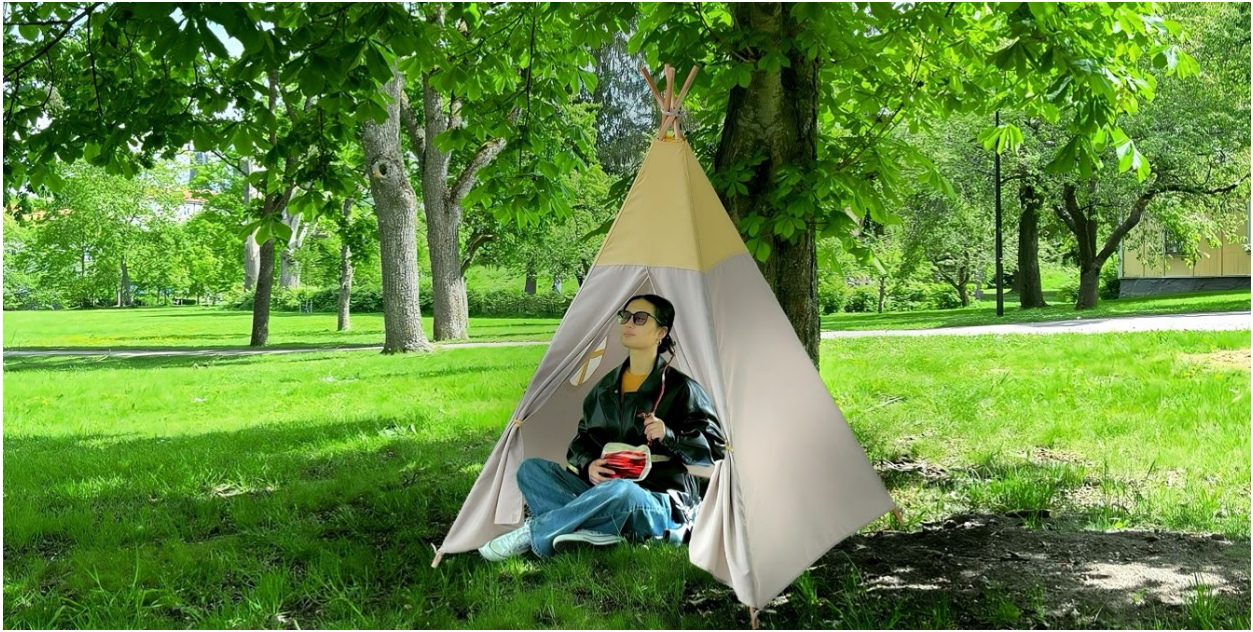

Figure 2: Outdoor demonstration of Umbilink, participants preparing to enter the enclosed space created by the tent

**Metaphor**

Umbilink uses the connection between the uterus and umbilical cord as its core symbolic structure, creating a tangible innovative interactive space. The healing space symbolises the womb, offering 'protection from the external world' and 'reconstruction of internal order'; the umbilical cord structure connects the user's body to the top of the space, evoking an emotional association of 'still being connected to the mother'; touch sensors are placed above the navel, activating the user's body memory, and the action of 'touching the navel' evokes psychological comfort and a sense of security. Through this, we have constructed a device system that uses bodily interaction to evoke emotional resonance, aiming to guide users to temporarily return to a primal sense of connection and calm amidst the complexities of reality.

## 4. METHODOLOGY

To obtain participants' subjective experiences when using the umbilical cord interface, this study followed a micro-phenomenological approach and adopted Petitmengin's elicitation interview as the data collection method [19]. In the semi-structured interviews, we encouraged participants to express themselves in the first person to help them become aware of their subtle subjective experiences when using Umbilink. Additionally, we encouraged participants to freely elaborate on their subjective narratives after answering the questions. Our participants were six HCI students (average age approximately 25 years old), including three females and three males. None of the participants had prior experience with childbirth. We conducted user testing using the standard prototype of Umbilink, with each participant's experience lasting approximately 10–15 minutes and the interview lasting approximately 15 minutes. Throughout this process, we recorded the interviews with the participants' informed consent. As part of a course project, team members also participated in testing activities for other groups, so participants did not receive compensation.

Following the grounded theory method, all three members of the team participated in the transcription and coding of the interview content, with the specific steps as follows: 1. **Data transcription:** All interview content was transcribed into text using speech recognition technology, with an average effective duration of approximately 12 minutes per person; 2. **Data cleaning:** We anonymised the six participants, replacing them with P1-P6; 3. **Triple Coding:** During the Open Coding phase, 25 initial tags such as 'transfer of control,' 'Buddhist memory,' 'wearable awkwardness,' and 'tactile distraction' were established; during the Axial Coding phase, we defined higher-level frameworks such as 'sensory reconstruction' and 'ritualism'; and in the final Selective Coding phase, we formed our core theoretical framework for umbilical cord interaction.

## 5. FINDINGS

**5.1 The Impact of Interpretation**

Our first users were a group of students closely connected to our team. Although we did not explain that the interactive device represented an electronic umbilical cord, they had a vague recollection of the concept of 'umbilical cord interaction' from previous seminars.

*'I did feel a sense of being wrapped up, perhaps due to psychological suggestion. I felt calm and peace... But, my feelings may be influenced by the introduction you gave me before this experience…You mentioned that you were interested in metaphors related to reproduction, so I assume this is simulating the umbilical cord and uterus? So, my feelings may be connected to your previous explanations...' (P3)*

*'During the experience, I tried to understand the metaphor related to the mother you mentioned earlier… Otherwise, none of us have seen an umbilical cord, and I wouldn't even associate it with this thing… Yes, if we knew nothing about it, this device would be quite strange, unless the environment is highly realistic…' (P4)*

The second user test took place after the prototype presentation was completed. We had previously detailed the conceptualisation process of this design. All users reported a sense of novelty during this informal experience and immediately imagined themselves as a foetus upon entering the device.

*"Can I lie inside it? This looks like something I could move to my dorm room—returning to my bed feels like returning to my mother's side… I love this idea; I bet I'd use it every day…" (Informal participant)*

*'After your presentation, I was wondering how the umbilical cord would be designed... I noticed you had carefully decorated the cables, though the texture might differ from a real umbilical cord, the appearance aligns with our understanding of it... So when I entered the space, I immediately immersed myself in the role—yes, I am a foetus, and in this corner, I truly felt joy...' (Informal participant)*

As an area that has received little attention and design consideration, the level of understanding of the umbilical cord significantly affects user experience and feedback. This was the most intuitive user feedback we received when we launched this interface prototype.

**5.2 Sensory Deprivation and Posture Maintenance**

All participants mentioned a certain 'failure' of the senses and body parts, along with the resulting sense of physical reset, with particular emphasis on the difference in the degree of the facial senses and body parts. This indicates that, despite our attempts to create a closed environment to simulate the state of the foetus in the womb, where the senses have not yet come into contact with the external environment. However, we overlooked the metaphorical significance of amniotic fluid, which serves as the physical environment directly experienced by the foetus, providing it with a 'weightless feeling.' This design oversight warrants further reflection.

*I like the all-black enclosed environment you created, rather than using many lights to decorate it… I want to feel my body, so any light would distract my attention… Besides ensuring that I'm not ganna fall, even though you continue your conversation, it's not my native language, and I feel like I don't understand English, as if all this has nothing to do with me… (P1)*

*'There was a moment when the environment was very quiet, and all my sensations focused on my navel. I felt its low-frequency vibrations and sound it emitted... I felt many of my sensations fading away, leaving only my brain... But my neck and legs still had clear sensations because I had to maintain my sitting posture'... (P5)*

**5.3 The Emergence of Cultural Memory**

Although Umbilink does not incorporate any cultural context to participants, several reported spontaneously triggering cultural memories during their experiences, using these to interpret and reshape the rhythmic nature of the installation. The non-prescriptive nature of the dark environment also served as a potential mediator for individual memories. The original intention of Umbilink was to suspend individual experiences, allowing users to return to primal state shared by all humans. However, participants' feedback still reflects the strong influence of individual cultural backgrounds on interface interpretation. Although a higher-fidelity prototype has not yet been tested, this indicates that we have a certain idealised tendency towards umbilical interaction.

*'While I felt a bit bored and couldn't help imagining myself in a temple in Sri Lanka... Maybe it's because i've been reading too many Buddhist books lately... I deliberately used some meditation techniques, like labelling distracting thoughts... After that, I almost tried to find distracting thoughts along with the frequency, and then the distracting thoughts disappeared...' (P3)*

*I recall that when I was a child, my mother would take me to Putuo Mountain every year to pray and make offerings, just like many people living in the Min Nan region, we have a close connection with Buddhism… Perhaps the frequency of the buzzer's sound is very low, somewhat like the sound of a wooden fish… But I hadn't thought about that scene in a long time…" (P2)*

**5.4 Ritual as a Mechnism**

At the beginning of our semi-structured interviews, several participants first reported their feelings during the process of wearing the device. In this overlooked stage, participants described how the wearing process influenced their experience, and how the sense of ritual and others' gaze became entangled with subsequent experiences in a vague manner.

*I think the most unique aspect was the process of putting it on. I had to pay attention to how to control the sensors. When I put on the waist pouch, I felt a sense of mystery and ritual… As if I had been given something, or perhaps I felt like a mother? I'm not sure… Suddenly, something extended from my waist, and it's hard to describe how that felt… It was definitely a new sensation… (P3)*

*"Let me put it this way: if I just sat down directly, then there would be auditory feedback from the environment, and it would be different from this… Even though everyone was watching me, putting on and taking off the device took some time, it was a bit awkward because it was really strange, but perhaps that's why it's so special…" (P6)*

## 6. DISCUSSION

In this section, we discuss the unique aspects of umbilical interaction design and the insights it offers. Drawing on the empirical knowledge gained from the 'Umbilical Interaction' practice, we first report on the findings from designing emerging interfaces under the metaphor of biological simulation. Secondly, we reflect on human-centred design principles. Based on the 'interface-centered design' scenario in this study, combined with Hookway's theory, we will discuss the ontology of human-computer interfaces, the fusion and separation of humans and systems, and their embodied interaction relationships. Finally, we will combine liminal rituals to discuss research gaps in human-computer interaction, such as device wearing and ritualistic elements, and their future applications.

### 6.1 Umbilical Interaction

Research on the umbilical cord as an embodied interface has yet to be explored by the academic community. This preliminary study provides pilot experience for future in-depth research in this direction. We have revealed the unique value of the umbilical cord as a human-computer interface. As a native material exchange interface for humans, it provides cognitive and cultural references for future flexible biomimetic design. As we anticipated during prototype design, participants criticised the vibration and sound feedback as insufficiently rich, while multimodal feedback including temperature changes was emphasised. Phenomenological research focused on specific modalities is a key priority for the next phase of research. The neural coding mechanisms of thermal and mechanical stimuli have been elucidated [20], and modal fusion strategies based on this foundation require further exploration. Our practice also refuted our unrecorded idealised assumption that umbilical interaction does not bring participants to a purely biological primal state. Even in a simulated environment of sensory degradation, participants still reconstruct their subjective experiences through cultural connections. In future umbilical interaction settings, personalised cultural imagination should not be ignored; instead, it should be incorporated into the introduction of metaphors, leveraging cultural consensus to eliminate intersubjectivity and thereby reduce cognitive load.

Additionally, previous embodied interfaces have emphasised reducing physiological load during interaction, often achieved through hand movements and whole-body activities [21]. However, in umbilical interactions simulating sensory deprivation, we found that while participants reported a certain degree of sensory deprivation, the effort to maintain posture has become a unique physiological load in this specific scenario. During development, the foetus achieves a relaxed body state by continuously resisting gravity through the buoyancy of amniotic fluid, a phenomenon referred to by Wilkinson et al. As the 'prenatal origin of anti-gravity homeostasis' [22]. Beyond the metaphorical correspondence between the maternal and the physical environment, amniotic fluid as a physical environment should be of noticed to future research, which provide a softer, more relaxed and close-body support environment for application in scenarios such as sleep, healing, and meditation.

### 6.2 Interface-Centered Design

Past HCI research advocate for the concealment and encapsulation of technology, such as hiding various services and protocols within network infrastructure to achieve seamless, human-centred design [23]. However, this study challenges the traditional design practice of the 'invisible interface' through Umbilink. We reveal how an explicit, stylised interface with distinct features can generate cognitive impact and emotional value. As defined by Hookway, interfaces function as abstract 'forms of relationship' [8] and require deliberate emphasis on their explicit presence to establish cognitive connections with users. In the current era where symbolism dominate information technology [24], the symbolic meaning inherent in embodied interfaces deserves greater attention. Interface-Centred Interaction provides a clear interpretive anchor for the cultural metaphors of interfaces.

The key difference between Umbilink and previous embodied interface designs lies in the fact that we did not attempt to enhance or extend any sensory functions. Instead, we sought to deprive participants of their subjective agency within the interactive device, in order to foster the integration of new body schemas. For RQ2, the shift of the 'the subject that to be served' toward the 'connecting end' is not merely a visual effect observed from a third-party perspective but rather the actual outcome of this degenerative simulation. During this process, participants reported sensations of sensory degradation and isolation from the external environment, along with a sense of tranquillity. This suggests that, to some extent, explicit embodied interfaces possss the potential to establish a new ontology of the 'human-interface-environment' trinity, contrasting with human-centred technological extensionism. Here, we also echo a post-human-centred design perspective, questioning the privileged status of human-centric values [25]. Although participants use Umbilink to gain influence over themselves, this process is achieved through sensory occlusion and being controlled. Drawing on Actor-Network Theory, the participants in this study, along with the occluded environment, electronic umbilical hardware, and other heterogeneous actors, form a dynamic interactive network through technological 'translation' [26]. In this post-human-centric framework, participants relinquish control to achieve novel embodied immersive experiences. Through the practice of Umbilink, we reflect on the questions raised by Hookway that the abstract nature of interfaces is inherent to their essence. When designers attempt to conceal the embodied nature, this often accompanies human-centred perspectives, including emerging interfaces such as BCI; conversely, when their existence is prominently displayed, the discomfort and strangeness evoke reflection on the existence of technology, thereby highlighting a post-human-centred perspective of human-machine equality.

**6.3 Liminality of Wearing**

Multiple participants reported that the peculiar sensation experienced during the process of wearing, the waist pouch significantly influenced their subjective perceptions. During this process, participants gradually detached from their original social role as 'researchers' within the former scenario and transitioned into a new role as "participants" , engaging in an unknown interactive experience. Victor Turner's limimal theory elucidates the 'ritualistic' transition reported by participants, where the wearing process appears to be a redundant and should be highly streamlined, is on the contrary, actually a liminal trigger in the embodied interaction process, which can promote Umbilink participants' exploration of their new selves [27]. As a cross-disciplinary theory widely applied in sociology, anthropology and other disciplines, liminal theory has applications in both costume design and human-computer interaction. For the former, clothing itself serves as a boundary between the dual roles of fashion designer and performer, has been deeply discussed [28]. The personalised value added in mass-produced clothing has also been compared by anthropologists to the wearing of adult rituals within liminal theory [29]. For the latter, different styles of virtual liminal spaces have been proven to have a crucial impact on interactive exploration [30], and the theory of liminal spaces in the real world has also been widely applied in the context of museums and other settings [31].

This study is based on the intersection of the two, providing a new perspective for HCI research: wearing process is a liminal ritual in embodied interaction. The unintentional complexity of the Umbilink wearing process is a violation of important qualities in structured design (such as usability or instant accessibility), yet it also embraces the unnoticed, 'invisible and transformative' phenomenological essence [32] of everyday life, which can be interpreted as a ritual of meaning-making rather than being guided by traditional evaluation metrics such as efficiency and usability. Additionally, the wearing ritual represents a shift in human-centric role positioning toward coexistence with technological devices. Our perspective aligns with Deroko's hybrid subjectivities approach and provides a new experimental field for emerging emotional dimensions, creating a liminal event model between humans and non-human technologies. [33]

**7. CONCLUSION**

Our research begins with a reflection on the traditional HCI research paradigm of Ubiquitous Computing and Human-Centred Design. Within the metaphor of the umbilical interaction space, we explore the possibility of returning to an 'interface-centric' interaction paradigm within specific contexts. In the Umbilink prototype, we designed a dark environment and an electronic umbilical cord providing rhythmic feedback, aiming to return participants to a primal state before sensory differentiation. We guided participants to describe their subjective experiences using first-person narratives, then analysed and summarised the feedback using grounded theory coding methods.

In this preliminary study, we questioned the 'seamless' quality of interaction design. For RQ1, we found that when the interface is significantly present between participants and the space, users actively perceive its existence and therefore understand it through its cultural attributes and metaphors. An unexpected finding was the ritualised wearing of embodied interfaces, which was considered a crucial factor influencing user experience in specific scenarios. We discussed this in conjunction with the theory of liminality. For RQ2, although users reported sensory degradation experiences when entering an immersive state, the absence of spatial support corresponding to the amniotic fluid environment, which required users to rely on there proprioception to paly additional effort to maintain posture. Additionally, participants suggested the need for multimodal and more precise feedback and reported a connection between umbilical interaction and there religious memories. As an absolutely emerging interface, designing with umbilical metaphor holds potential value for future exploration of topics related to healing, meditation, and sleep. The involvement of higher-fidelity prototypes and other research methods will be crucial for the next phase of research.